# Molecular Energy Relations From Chemical Kinetics

**Robert W. Finkel**

*Physics Department, St. John's University, Jamaica, New York 11439*

Since molecular energy transformations are responsible for chemical reaction rates at the most fundamental level, chemical kinetics should provide some information about molecular energies. This is the premise and objective of this note. We describe a Hamiltonian formulation for kinetic rate equations where the concentrations are the generalized coordinates and the conjugate momenta are simply related to individual average molecular energies. Simple examples are presented and the resulting energy relations naturally include non-equilibrium reactions. An analysis predicts the reasonable outcome that thermal agitation of a composite molecule increases its rate of dissociation.

## Introduction

Quantum considerations require that the rate coefficients of chemical kinetic equations be related to the interaction energies of the participating molecules. In practice, however, chemical kinetics and the corresponding energy relations are usually treated separately and independently. Exceptions to this dichotomy are seen in transition state theories[1] where rate coefficients are determined by theoretical expressions involving molecular energies. Relationships between reaction rate coefficients and thermodynamic quantities should exist even when both are phenomenological and separately determined.

A theory capable of predicting energy relations from chemical kinetics was formulated in a quantum mechanical context,[2] but a feature that describes individual molecular energies was not utilized. The theory is based on a Hamiltonian function constructed from an arbitrary set of rate equations. A classical version of this formalism[3] enabled rate equations to be treated with the techniques of canonical transformations, perturbation theory, and dynamic invariants. Again, the feature sought here– prediction of individual molecular energies from chemical kinetic equations–was not used. This energy feature was recently applied to a problem of active transport across membranes.[4] However, no development of the theory or methodology was included and that note should properly be regarded as subsequent to the present one.

In the next section we summarize the Hamiltonian formulation that was originally developed for a quantum problem in reference 2. Examples and interpretations follow to reveal some practical and theoretical features of the methodology. Finally, we show that the quantum genesis of the theory is spurious by reproducing the results with a classical derivation.

## Hamiltonian Formulation

Consider a uniform, isothermal solution in a unit volume of $m$ chemical species with concentrations $q_1 \ldots q_m$. These will serve as generalized coordinates in a Hamiltonian. The time development is governed by a set of $m$ autonomous rate equations,

$$\dot{q}_j = f_j(q_1 \cdots q_m)$$

where a dot denotes a time derivative and $j$ runs from 1 to $m$. The Hamiltonian $H$ is then simply constructed,

$$H = \sum_j [f_j p_j + \mu_j(q_j) q_j], \qquad (1)$$

where $p_j$ is a canonical momentum corresponding to $q_j$ and $\mu_j(q_j)$ is the associated residual energy independent of interactions with the other chemical species. The Hamiltonian of equation (1) is degenerate in that it cannot be transformed to a Lagrangian by a Legendre transform. This does not hamper its utility.

The physical interpretation of the action $p_j$ is our particular concern here. Reference 2 established that $-dp_j/dt$ is the individual particle energy $\varepsilon_j$ of the $j$th species. The per-particle energy then follows from the Hamilton equation $dp_j/dt = -\partial H/\partial q_j$,

$$\varepsilon_j = \partial H/\partial q_j. \qquad (2)$$

## Example

A simple example will serve to illustrate the formalism. According to equation (1), the Hamiltonian for the second order reaction,

$$A + B \xrightarrow{k} C \quad (3)$$

is

$$H = k\,a\,bP + H_0 \quad (4)$$

where the lower case letters signify concentrations or particle numbers and

$$P \equiv p_C - p_A - p_B$$
$$H_0 \equiv a\mu_A + b\mu_B + c\mu_C$$

The $\mu$'s are chemical potentials having the ideal forms $\mu_A = k_B T \ln(a) + \mu_A^0$ with Boltzmann's constant $k_B$, temperature $T$, and a "standard" energy $\mu_A^0$.

Applying equation (2) gives energies in terms of $P$,

$$\varepsilon_A = kbP + \mu_A + k_B T$$
$$\varepsilon_B = kaP + \mu_B + k_B T$$
$$\varepsilon_C = \mu_C + k_B T$$

As a practical matter, we must evaluate $P$. This can be calculated from the differential equations generated by Hamilton's equations, but it is simpler to use an energy expression involving $P$.

It can be seen that the terms containing $P$ express the mutual molecular interactions. Assuming $a < b$ and equating the energies of reactants and products, we write

$$\varepsilon_A + \mu_B + k_B T = \varepsilon_C.$$

The rationale is that $\varepsilon_A$ includes the interaction energy of $A$ with $B$ and adding the "rest energy" of $B$ produces the total energy of reactants. Solving for $P$ and substituting into the energy expressions gives

$$\varepsilon_A = k_B T \ln K + \mu_A$$
$$\varepsilon_B = (a/b)(k_B T \ln K - k_B T) + \mu_A + k_B T$$
$$\varepsilon_C = \mu_C + k_B T$$
$$K \equiv \frac{c}{ab}.$$

Some particulars can be addressed in the context of this example. $K$ has the formal appearance of an equilibrium constant, but the system is clearly not in equilibrium and the molecular interaction energies are manifestly concentration-dependent. Another feature to note is the appearance of $k_B T$ as part of the intrinsic molecular energies. This has been shown to be the average thermal interaction energy for a molecule with its environment.[5] It appears naturally in this formalism.

## Physical Interpretation

A demonstration for arbitrary second order rate equations provides a somewhat more physical view of the formalism. Consider second order rate equations of the form

$$f_j = \tfrac{1}{2}\sum_{i,k} C_{ik}^j q_i q_k + \sum_i C_i^j q_i$$

where the $C$'s are constant coefficients and $C_{ik}^j = C_{ki}^j$. Substituting the rates into the Hamiltonian (1) and applying equation (2) produces energies with the form

$$\varepsilon_j = \sum_i V_{ji} q_i + U_j$$

where $V_{ji}$ and $U_j$ linear functions of $p$'s. Since $\varepsilon_j$ is the average energy of a $j$-molecule, we can identify the coefficient $V_{ji}$ as the average two-particle interaction energy between species $i$ and $j$. Then $U_j$ represents all energies exclusive of pair interactions including internal potential energy, average kinetic energy, and average energy of interaction with the solution or an external source.

## Reversible Reaction

An application to a reversible reaction describes energetics away from the equilibrium point and reduces to familiar expressions at equilibrium. A Hamiltonian for the reaction $A + B \Leftrightarrow C$ is

$$H = (k_{-1}c - k_1 ab)P + H_0$$

where the rate coefficients for the forward and reverse reactions are respectively $k_1$ and $k_{-1}$ and $P$ and $H_0$ are defined as for equation (4).

We apply equation (2) and, as before, assume $a < b$ to equate the energies of reactants and products, $\varepsilon_A + \mu_B + k_B T = \varepsilon_C.$ Now solving for $P$ and substituting into the energy expressions gives

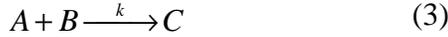

$$\varepsilon_A = -\frac{k_1 b}{k_1 b + k_{-1}}(\Delta G + k_B T) + \mu_A + k_B T$$

$$\varepsilon_B = -\frac{k_1 a}{k_1 b + k_{-1}}(\Delta G + k_B T) + \mu_B + k_B T$$

$$\varepsilon_C = \frac{k_{-1}}{k_1 b + k_{-1}}(\Delta G + k_B T) + \mu_C + k_B T$$

$$\Delta G \equiv -k_B T \ln K, \text{ with } K \cong c/(ab)$$

The results are informative even for this prosaic reaction. The notation $\Delta G$ seems appropriate for the above quantity, but we emphasize that it does *not* correspond to Gibbs free energy except at the equilibrium point. The results shown here are valid away from equilibrium. A negative interaction energy is associated with a favored state and we see that for positive $\Delta G$ the reactants $A$ and $B$ are favored. Conversely, the product $C$ then has a positive interaction energy and its dissociation is favored as expected.

An unanticipated outcome occurs when our $\Delta G$ vanishes and the reactants are still favored slightly over the product. This differs from standard thermodynamics where $\Delta G = 0$ signifies equilibrium for the reaction.[6] We attribute this difference to the fact that the solution or environment imparts a thermal assault that slightly destabilizes the $C$ composite. This factor is reasonably neglected in thermodynamic treatments. Reaction equilibrium in the present case requires $\Delta G = -k_B T$ and the forward reaction is spontaneous when $\Delta G < -k_B T$. Although the effect is small, an experimental test would have implications for the current theory.

## Classical Derivation

It is clear that all quantum attributes are subsumed in kinetic coefficients and energy parameters so we have an essentially classical theory. Consequently, it is desirable to reestablish the Hamiltonian approach for chemical kinetics in a classical format.

We can, in principle, treat a homogenous system of interacting particles as if each particle is an isolated entity in a pseudo potential. In notation familiar in Hamilton-Jacobi theory,[7,8] let $q_i$ represent the number of species $i$ molecules and let $s_i$ represent Hamilton's principal function for an individual molecule at its center of mass. We now construct a principal function $S$ that describes the aggregate system,

$$S = \sum_i q_i s_i. \quad (5)$$

The system so characterized is *not* the original system. Rather, it treats each species as if it is a separate and independent system. Mutual interaction energies may overlap or cancel.

Take the time derivative of equation (5),

$$\dot{S} = \sum \dot{q}_i s_i + \sum q_i \dot{s}_i \quad (6)$$

Consider $s_i$ expressed in terms of its local position coordinates, $x_k$,

$$ds_i/dt = \partial s_i/\partial t + \sum_k \dot{x}_k \, \partial s_i/\partial t.$$

Velocity vanishes for a particle in its center of mass frame so the last term vanishes and the partial derivative of $s_i$ can be replaced by its total derivative. From Hamilton-Jacobi theory, the energy of the individual molecule of species $i$, $\varepsilon_i$, is given by $\varepsilon_i = -\partial s_i/\partial t$. By construction, $\dot{S}$ can be expressed as a function of $q$'s and we write $\dot{S} \equiv -H_0$. Substituting into equation (6), we have

$$-H_0 = \sum \dot{q}_i s_i - \sum q_i \varepsilon_i \quad (7)$$

The last term is the total system energy and we identify it with a Hamiltonian,

$$H = \sum \dot{q}_i s_i + H_0 \quad (8)$$

and comparing the relation $\partial H/\partial s_i = \dot{q}$ with Hamilton's equation for $\dot{q}$ identifies $s_i$ as the conjugate momentum to $q_i$. With this identification, we restore our original notation, $p_j \equiv s_j$.

## Discussion

This note focused on reintroducing the energetic aspects of the Hamiltonian theory of chemical kinetics in a classical context. Although the theory was exemplified with only the simplest second order reactions, some notable characteristics were evident.

The most obvious feature of energy applications is that they apply generally to non-equilibrium processes. The familiar form for Gibb's free energy, $\Delta G$, arises naturally from the theory and is not limited to equilibrium reactions. The sign of the computed interaction energy decides the direction of spontaneous reaction and this is in substantial agreement with equilibrium thermodynamic requirements on the sign of ΔG. It was recognized that thermal agitation of a composite molecule increases its destabilization; this reasonable prediction resolves a minor difference with the standard thermodynamics treatment and follows as an inherent consequence and prediction of the current theory.

[1] S. Glasstone, K. J. Laidler, and H. Eyring, *The Theory of Rate Processes: the Kinetics of Chemical Reactions, Viscosity, Diffusion and Electrochemical Phenomena* (McGraw-Hill, New York, 1941)

[2] R. W. Finkel, Nuovo Cimento **44B**, 135 (1978).

[3] R. W. Finkel, J. Phys. Chem. A **105**, 9219 (2001).

[4] R. W. Finkel, arXir: q-bio SC/0702026

[5] R. W. Finkel and J. Joseph, Nuovo Cimento **79B**, 155 (1984).

[6] R. P. Bauman, *Modern Thermodynamics with Statistical Mechanics* (Macmillan, New York, 1992), Chap. 11, p.222.

[7] H. Goldstein, *Classical Mechanics*, **2nd** edition (Addison-Wesley, Reading, MA, 1980)

[8] A. L. Fetter and J. D. Walecka, *Theoretical Mechanics of Particles and Continua* (Dover, Mineola, New York, 2003)